\documentclass{article}
\usepackage[T1]{fontenc} 
\usepackage[utf8]{inputenc} 
\usepackage{ismir,amsmath,cite,url}
\usepackage{graphicx}
\usepackage{color}
\usepackage{comment}
\usepackage[numbers, sort&compress, square]{natbib}
\usepackage{amsmath}
\usepackage{amssymb}
\usepackage{amsfonts}
\usepackage{amsthm}
\usepackage{bm}
\usepackage{mathrsfs}

\usepackage{siunitx}

\usepackage{enumitem}

\usepackage{placeins}
\usepackage{graphicx}

\usepackage{booktabs}
\usepackage{tabularx}
\usepackage{xcolor}
\usepackage{colortbl}
\usepackage{arydshln}

\usepackage{multicol}
\usepackage{mdframed}
\usepackage{xspace}

\def\lbd{}	

\usepackage{lineno}

\usepackage{lipsum}

\usepackage[skip=0pt]{caption}

\usepackage{jabbrv}

\DefineJournalException{Extended Abstracts of the ICML Workshop on Machine Learning for Music Discovery}{Extended Abstracts for ML4MD, ICML}

\DefineJournalException{Proceedings of the 34th International Conference on Machine Learning}{Proc. ICML}
\DefineJournalException{Proceedings of the 35th International Conference on Machine Learning}{Proc. ICML}
\DefineJournalException{Proceedings of the 36th International Conference on Machine Learning}{Proc. ICML}

\DefineJournalException{Proceedings of the 2nd International Conference on Learning Representations}{Proc. ICLR}
\DefineJournalException{Proceedings of the 5th International Conference on Learning Representations}{Proc. ICLR}
\DefineJournalException{Proceedings of the 6th International Conference on Learning Representations}{Proc. ICLR}
\DefineJournalException{Proceedings of the 7th International Conference on Learning Representations}{Proc. ICLR}
\DefineJournalException{Proceedings of the 8th International Conference on Learning Representations}{Proc. ICLR}

\DefineJournalException{Proceedings of the 15th International Conference on Neural Information Processing Systems}{Proc. NeurIPS}
\DefineJournalException{Proceedings of the 30th International Conference on Neural Information Processing Systems}{Proc. NeurIPS}
\DefineJournalException{Proceedings of the 32nd International Conference on Neural Information Processing Systems}{Proc. NeurIPS}

\DefineJournalException{Proceedings of the 28th International Joint Conference on Artificial Intelligence}{Proc. IJCAI}

\DefineJournalException{Proceedings of the 19th International Society for Music Information Retrieval Conference}{Proc. ISMIR}
\DefineJournalException{Proceedings of the 20th International Society for Music Information Retrieval Conference}{Proc. ISMIR}
\DefineJournalException{Proceedings of the 21st International Society for Music Information Retrieval Conference}{Proc. ISMIR}
\DefineJournalException{Proceedings of the 22nd International Society for Music Information Retrieval Conference}{Proc. ISMIR}
\DefineJournalException{Proceedings of the 22nd International Society for Music Information Retrieval Conference}{Proc. ISMIR}

\title{%
    Evaluation of Latent Space Disentanglement\\in the Presence of Interdependent Attributes%
}

\multauthor
{%
    Karn N. Watcharasupat$^{1, 2}$ \hspace{1cm} 
    Alexander Lerch$^1$
}{%
    $^1$Center for Music Technology, Georgia Institute of Technology, Atlanta, GA, USA\\
    $^2$School of Electrical and Electronic Engineering, Nanyang Technological University, Singapore\\
    {\tt%
        karn001@e.ntu.edu.sg,
        alexander.lerch@gatech.edu
    }
}

\def\authorname{%
    K. N. Watcharasupat, 
    and A. Lerch
}

\usepackage[bookmarks=false,pdfauthor={\authorname},pdfsubject={\papersubject},hidelinks]{hyperref}
\usepackage{cleveref}

\usepackage{etoolbox}
\AtBeginEnvironment{align}{%
  \linenomath
}
\AtEndEnvironment{align}{%
  \endlinenomath
}

\begin{document}
\maketitle

\begin{abstract}
Controllable music generation with deep generative models has become increasingly reliant on disentanglement learning techniques. However, current disentanglement metrics, such as mutual information gap (MIG), are often inadequate and misleading when used for evaluating latent representations in the presence of interdependent semantic attributes often encountered in real-world music datasets. In this work, we propose a dependency-aware information metric as a drop-in replacement for MIG that accounts for the inherent relationship between semantic attributes.
\end{abstract}

\begin{figure*}[!t]
    \centering
    \includegraphics[width=0.9\textwidth]{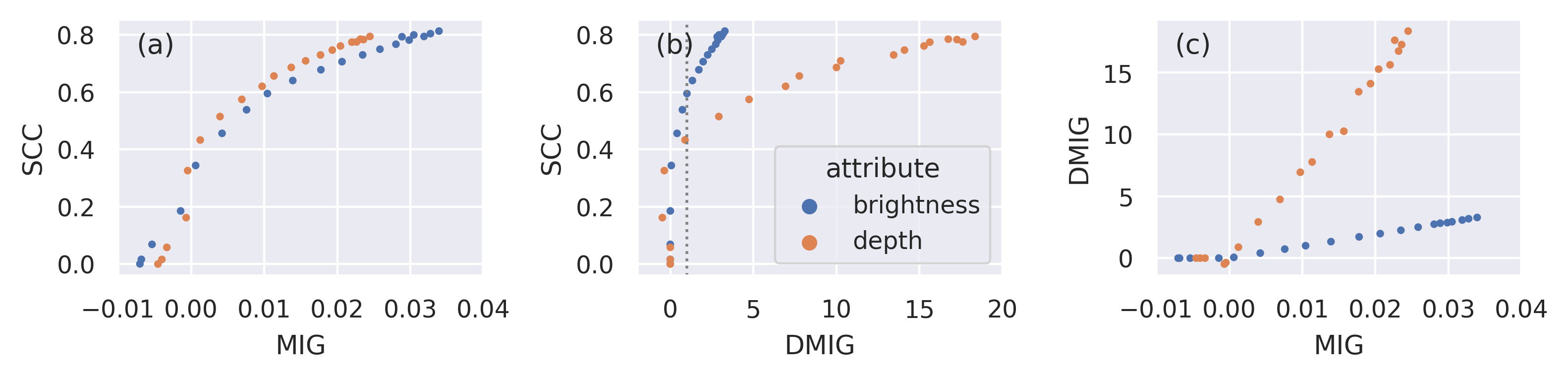}
    \caption{Plots of (a) SCC against MIG, (b) SCC against DMIG, and (c) DMIG against MIG, on the validation set.}
    \label{fig:plot}
\end{figure*}

\section{Introduction}

Disentanglement learning has been an influential field of studies for controllable music generation with variational autoencoders (VAEs). A number of previous studies have attempted supervised disentanglement learning techniques on several semantic attributes such as rhythm, pitch range \cite{Pati2019LatentAttributes}, note density, contour \cite{Pati2020Attribute-basedAuto-encoders}, arousal \cite{Tan2020MusicModelling}, style \cite{Hung2019MusicalRepresentations}, and genre \cite{Brunner2018MIDI-VAE:Transfer} to varying degrees of success. However, learning to simultaneously manipulate multiple attributes, in particular, remains a difficult task to both achieve and objectively evaluate \cite{Pati2021IsGeneration} due to the limitation of current metrics. 

One major issue with popular disentanglement metrics \cite{Do2020TheoryRepresentations}, such as mutual information gap (MIG) \cite{Chen2018IsolatingAutoencoders}, separate attribute predictability (SAP) \cite{Kumar2018VariationalObservations}, and modularity \cite{Ridgeway2018LearningLoss}, is that they were designed for independent generative factors, rather than real-world semantic attributes. As semantic attributes related to music are often highly interdependent, these metrics do not provide an accurate reflection of the `quality' of learnt latent representation regularized for multiple interdependent attributes. Information inherently shared between attributes is penalized in the same way as that due to undesired entanglement issues.

In this work, we propose a dependency-aware metric based on mutual information (MI) to act as a drop-in replacement for MIG. Preliminary experiments were carried out to demonstrate the benefits of the proposed metrics over MIG.

\section{Proposed Metrics}

Consider a set of attributes $\{a_i\}_{i=1}^{M}$ and a latent vector $\bm{z}\in\mathbb{R}^D$ with $M\le D$. Without loss of generality, for $i \le M$, we assume $z_i$ is regularized for $a_i$. The remaining dimensions are unregularized. $\mathcal{H}(\cdot)$ denotes entropy while $\mathcal{I}(\cdot, \cdot)$ denotes mutual information.


MIG was proposed in \cite{Chen2018IsolatingAutoencoders} to measure the degree of disentanglement in a latent space. The idea behind MIG can be said to measure: \textit{for each attribute, the normalized difference between the mutual information between the attribute and its most informative latent dimension, and that between the attribute the second-most informative latent dimension}.  Mathematically, MIG is given by
\begin{equation}
    \text{MIG}(a_i) = 
    \left(\mathcal{I}(a_i, z_i) -  \mathcal{I}(a_i,z_j)\right)/\mathcal{H}(a_i),
\end{equation}
where $j=\arg\max_{k \ne i} \mathcal{I}(a_i,z_k)$. It is reasonable to assume $i=\arg\max_{k} \mathcal{I}(a_i,z_k)$ in a supervised setting; otherwise MIG takes negative values to indicate regularization failure. The normalization is given by $\mathcal{H}(a_i)$, which would be the maximum possible difference in MI between a latent dimension $z_i$ coding perfectly for $a_i$, i.e., $\mathcal{I}(a_i, z_i) = \mathcal{H}(a_i)$ and the second-most containing no information about $a_i$, i.e., $\mathcal{I}(a_i, z_j) = 0$. As such, MIG is bounded above by one.


However, given the interdependence of semantic attributes, if $j \le M$, the ideal value of the difference $\mathcal{I}(a_i, z_i) -  \mathcal{I}(a_i,z_j)$ is no longer $\mathcal{H}(a_i)$ since
\begin{equation}
    \mathcal{I}(z_j, a_j) > 0 \wedge \mathcal{I}(a_i, a_j) > 0 \implies \mathcal{I}(a_i, z_j) > 0.
\end{equation}
For regularized latent dimensions, we consider a pair of inherently entangled attributes $(a_i, a_j)$, i.e., $\mathcal{I}(a_i, a_j) > 0$. Under the ideal case where $z_i$ is fully informative \cite{Do2020TheoryRepresentations} about $a_i$, i.e.,  $\mathcal{H}(a_i|z_i) = 0$, we have
\begin{align}
    &\mathcal{I}(a_i, z_i) - \mathcal{I}(a_i,z_j)\nonumber\\
    &\qquad= \left[\mathcal{H}(a_i) - \mathcal{H}(a_i|z_i)\right]-\left[\mathcal{H}(a_i) - \mathcal{H}(a_i|z_j)\right] \\
    &\qquad= \mathcal{H}(a_i|z_j) \qquad \because \mathcal{H}(a_i|z_i) = 0.
\end{align}
    
Moreover, in the ideal case, $z_j$ and $a_j$ also have an invertible mapping between each other, this means that $\mathcal{H}(a_i|z_j) = \mathcal{H}(a_i|a_j)$. Hence, in the ideal case, the difference is given by
\begin{equation}
\mathcal{I}(a_i, z_i) - \mathcal{I}(a_i,z_j)= \mathcal{H}(a_i|a_j).
\end{equation}
As such, we extend the definition of mutual information gap to the dependency-aware mutual information gap (DMIG) as follows
\begin{equation}
    \text{DMIG}(a_i) = 
    \begin{cases}
    \left(\mathcal{I}(a_i, z_i) -  \mathcal{I}(a_i,z_j)\right)/\mathcal{H}(a_i|a_j) & j \le M\\
     \left(\mathcal{I}(a_i, z_i) -  \mathcal{I}(a_i,z_j)\right)/\mathcal{H}(a_i) & j > M,
    \end{cases}
\end{equation}
where $j=\arg\max_{k \ne i} \mathcal{I}(a_i,z_k)$. DMIG remains faithful to the core idea of MIG but modifies the normalization to properly account for inter-attribute dependencies. When $a_i$ and $a_j$ are independent, $\mathcal{H}(a_i) - \mathcal{I}(a_i, a_j) =\mathcal{H}(a_i)$ and the DMIG reduces to vanilla MIG. 

Note that in the case of continuous random variables, differential entropy can be negative, unlike discrete Shannon entropy. This is particularly evident with conditional differential entropy and may result in DMIG values above unity whenever $\mathcal{H}(a_i|z_i)/\mathcal{H}(a_i|z_j)$ is negative.

\section{Experiments}

To illustrate the key features of the dependency-aware metrics, we evaluate the latent space of a VAE model trained to reconstruct raw musical audio while being regularized for two highly correlated attributes\footnote{See the supplementary materials for full experimental details at \href{https://github.com/karnwatcharasupat/dependency-aware-mi-metrics}{https://github.com/karnwatcharasupat/dependency-aware-mi-metrics}.}.

\subsection{Data and model}

We use the NSynth dataset \cite{Engel2017NeuralAutoencoders}, which is a large-scale dataset of musical notes played by various instruments with diverse timbral qualities. The dataset provides 4-second snippets sampled at \SI{16}{\kilo\hertz}. From the raw audio provided by NSynth, we extract two semantic attributes, namely, \textit{brightness} and \textit{depth} using the AudioCommons Timbral Model \cite{Pearce2019DeliverableContent}. Since both the brightness and depth features are heavily influenced by the spectral distribution of the sound \cite{Pearce2017DeliverableContent}, they are strongly correlated. 

We trained a convolutional VAE model to reconstruct the log-magnitude spectrogram of the audio and obtain reconstructed 
time-domain audio using a phase-bypass reconstruction. The models are trained using the attribute-regularized $\beta$-VAE loss function \cite{Kingma2014Auto-encodingBayes, Higgins2017, Pati2020Attribute-basedAuto-encoders}
\begin{equation}
    \mathcal{L} = \mathcal{R}(\hat{\mathbf{x}}; \mathbf{x}) + \beta \mathcal{D}\left(\mathbf{z}\right)
    + \gamma \textstyle\sum_{i}\mathcal{A}(z_i; a_i),
\end{equation}
where $\mathcal{R}(\cdot)$ is the reconstruction loss implemented via the mean square error on the log-magnitude spectrograms, $\mathcal{D}(\cdot)$ is the KL divergence term with a standard normal prior, and $\mathcal{A}(\cdot)$ is the AR-VAE regularization from \cite{Pati2020Attribute-basedAuto-encoders}.
We used $D=\num{512}$, $\beta=\num{1}$, and $\gamma=\num{10}$.

\subsection{Results}

\Cref{fig:plot} plots the MIG, DMIG, and Spearman correlation coefficient (SCC) of the attributes (brightness and depth) with respect to their respective regularized latent dimensions on the validation set over the course of the training. Due to the high correlation between brightness and depth, for most of the training, the most and second-most informative latent dimensions in MIG/DMIG are the regularized ones that encode for the attributes.

As seen from \Cref{fig:plot}(a), the MIG values are generally very low (in the order of \num{e-2}, out of maximum 1) despite the SCC indicating successful encoding of the attribute information into the latent dimension. This is due to the high mutual information between brightness and depth, resulting in a very low true bound for MIG. On the other hand, we can observe from \Cref{fig:plot}(b) that DMIG reflects more clearly the quality of the latent space as it encodes the attribute; the rapid improvement in SCC mostly occurred before DMIG reaches one (dotted line). In \Cref{fig:plot}(c), the highly linear relationship between MIG and DMIG further demonstrates the idea that DMIG is simply MIG renormalized to better reflect the dependencies between semantic attributes coded by the model. Admittedly, the peculiarities of differential conditional entropy and the practical computation of mutual information and entropy estimates \cite{scikit-learn} have contributed to a DMIG range that is much larger than vanilla MIG. We will be working to resolve this limitation in future work. 

\section{Conclusion}

In this work, we propose a dependency-aware extension to a popular disentanglement metrics, mutual information gap (MIG), to better account for inter-attribute dependencies often observed in real-world datasets. Key features of the proposed dependency-aware MIG were demonstrated via an experiment on an audio dataset with highly correlated timbral attributes.

\section{Acknowledgement}
K. N. Watcharasupat acknowledges the support from the CN Yang Scholars Programme, Nanyang Technological University, Singapore.

\setlength{\bibsep}{0pt}
\renewcommand{\bibsection}{\section{References}}
\bibliography{references}

\end{document}